\documentclass[conference]{IEEEtran}
\IEEEoverridecommandlockouts

\usepackage{cite}
\usepackage{amsmath,amssymb,amsfonts}
\usepackage{algorithmic}
\usepackage{graphicx}
\usepackage{textcomp}
\usepackage{xcolor}

\usepackage{amsmath,graphicx}
\usepackage{booktabs}
\usepackage{tabularx}
\usepackage{multirow}
\usepackage{mathtools}
\usepackage{arydshln}
\usepackage{gensymb}
\usepackage{textcomp}
\usepackage{amsfonts}
\usepackage{amsthm}
\usepackage{eucal}
\usepackage{microtype}
\usepackage{hyperref}       
\usepackage{url}            
\usepackage{nicefrac}       
\usepackage{subfigure}
\usepackage{xcolor}
\usepackage{colortbl}
\usepackage{enumitem}

\usepackage{tabularray}
\NewTableCommand\SCC[1]{\SetCell{bg=#1}}

\def\BibTeX{{\rm B\kern-.05em{\sc i\kern-.025em b}\kern-.08em
    T\kern-.1667em\lower.7ex\hbox{E}\kern-.125emX}}
\begin{document}



\title{MECG-E: Mamba-based ECG Enhancer for Baseline Wander Removal}

\author{
\IEEEauthorblockN{Kuo-Hsuan Hung\IEEEauthorrefmark{1}, Kuan-Chen Wang\IEEEauthorrefmark{2}, Kai-Chun Liu\IEEEauthorrefmark{3}, Wei-Lun Chen\IEEEauthorrefmark{2}, 
Xugang Lu\IEEEauthorrefmark{4}\\ Yu Tsao\IEEEauthorrefmark{2}
and Chii-Wann Lin\IEEEauthorrefmark{1}} 
\IEEEauthorblockA{\IEEEauthorrefmark{1}\textit{Department of Biomedical Engineering, National Taiwan University, Taiwan}}
\IEEEauthorblockA{\IEEEauthorrefmark{2}\textit{Research Center for Information Technology Innovation, Academic Sinica, Taiwan}}
\IEEEauthorblockA{\IEEEauthorrefmark{3}\textit{College of Information and Computer Sciences, University of Massachusetts Amherst, USA}}
\IEEEauthorblockA{\IEEEauthorrefmark{4}\textit{National Institute of Information and Communications Technology, Japan}}
\IEEEauthorrefmark{1}\textit{\{d07528023, cwlinx\}@ntu.edu.tw}, \IEEEauthorrefmark{2}\textit{d12942016@ntu.edu.tw}, 
\IEEEauthorrefmark{3}\textit{kaichunliu@umass.edu}\\ 
\IEEEauthorrefmark{2}\textit{\{rogetlio1121, yu.tsao\}@citi.sinica.edu.tw}, \IEEEauthorrefmark{4}\textit{xugang.lu@nict.go.jp} 
}

\maketitle

\begin{abstract}
Electrocardiogram (ECG) is an important non-invasive method for diagnosing cardiovascular disease. However, ECG signals are susceptible to noise contamination, such as electrical interference or signal wandering, which reduces diagnostic accuracy. Various ECG denoising methods have been proposed, but most existing methods yield suboptimal performance under very noisy conditions or require several steps during inference, leading to latency during online processing. In this paper, we propose a novel ECG denoising model, namely Mamba-based ECG Enhancer (MECG-E), which leverages the Mamba architecture known for its fast inference and outstanding nonlinear mapping capabilities. Experimental results indicate that MECG-E surpasses several well-known existing models across multiple metrics under different noise conditions. Additionally, MECG-E requires less inference time than state-of-the-art diffusion-based ECG denoisers, demonstrating the model's functionality and efficiency.
\end{abstract}



\section{Introduction}
Cardiovascular diseases are the leading cause of death worldwide\footnote{https://www.who.int/health-topics/cardiovascular-diseases}, necessitating the accurate and reliable diagnostic tools. Among these, the electrocardiogram (ECG) plays a crucial role  as a  non-invasive method for monitoring the heart’s electrical activity \cite{dai2021convolutional}. However, ECG signal acquisition is frequently compromised by various types of noise and artifacts, which can significantly degrade signal quality and impact diagnostic accuracy \cite{hu2024lightweight}.

To address the above issues, several signal processing techniques have been proposed. Commonly used filter-based methods include Finite Impulse Response (FIR) filters \cite{kumar2015removal, van1985removal}, Infinite Impulse Response (IIR) filters \cite{pottala1990suppression}, and adaptive Kalman filters \cite{vullings2010adaptive}. Additionally, mode-decomposition-based approaches \cite{huang1998empirical, blanco2008ecg, kabir2012denoising} has been employed to decompose noisy signals into intrinsic mode functions (IMFs), with the IMFs containing the most noise being removed. Statistical methods, such as Independent Component Analysis (ICA) \cite{barati2006baseline, he2006application}, have also been explored. Among transform-based methods, wavelet transform (WT) techniques \cite{sayadi2007multiadaptive, reddy2009ecg, liu2013novel} have been primarily developed and have shown promising results. Despite their widespread application, these traditional methods are typically effective under ideal conditions (e.g., high signal-to-noise ratio and single noise type), and often heavily rely on human expertise or trial-and-error.  These limitations and challenges notably constrain their denoising ability
and generalizability in challenging scenarios.


In recent years, the advent of deep learning (DL) has revolutionized the field of ECG signal processing. Various deep learning methods have been introduced to improve the quality of ECG signals by effectively eliminating noise and artifacts, such as deep recurrent neural networks (DRNN) \cite{antczak2018deep} and fully convolutional networks (FCN) \cite{chiang2019noise}. Additionally, DeepFilter \cite{romero2021deepfilter} has introduced Multi-Kernel Linear And Non-Linear (MKLANL) filter modules for multi-scale feature denoising. The use of Generative Adversarial Networks-based model (GANs) \cite{singh2020new, wang2022ecg} to generate high-fidelity ECG signals from noisy inputs has also been explored. 

\begin{figure*}[!htb]
\centerline{\includegraphics[width=1.9\columnwidth]{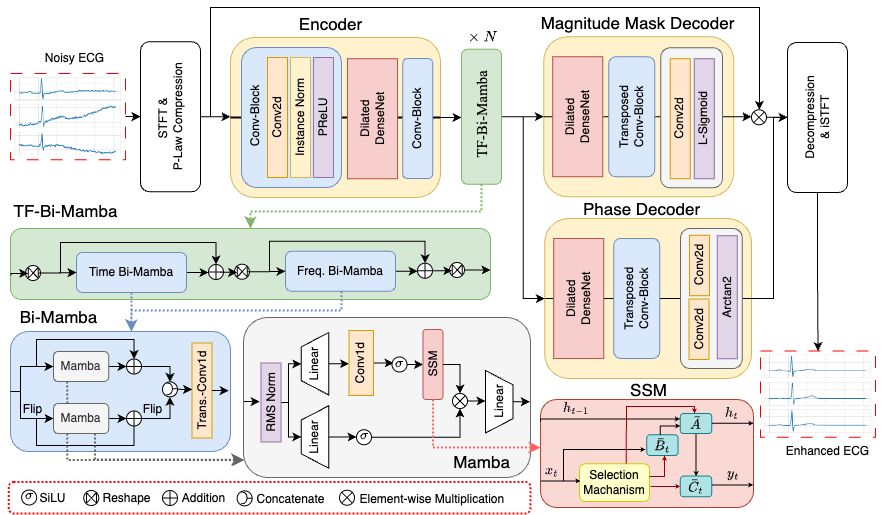}}
\caption{The overview of the proposed MECG-E model. The dashed line represents the corresponding module. The SSM modules refer to the selective state space mechanism mentioned in Sec.\ref{sec:mamba}.
}
\label{fig:ECGSE}
\end{figure*}

Recently, with the growing interest in diffusion models, score-based diffusion models \cite{li2023descod,el2024wavelet} have been applied and have notably outperformed other methods. However, diffusion models require numerous sampling steps to achieve the desired quality, resulting in considerably slower inference speeds. 

To address the challenges of suboptimal performance and slow inference speeds in existing ECG denoising methods, we propose a novel model, Mamba-based ECG Enhancer (MECG-E). MECG-E incorporates the Mamba framework, which is known for its rapid inference and high performance \cite{gu2023mamba}, into an advanced ECG signal enhancement architecture. The key contributions of this paper are outlined as follows:
\begin{itemize}
\item MECG-E consistently outperforms existing models across various metrics under a wide range of noise levels. Compared to the state-of-the-art (SOTA) model, it significantly reduces the inference time, proving its efficiency and effectiveness in practical applications.
\item To the best of our knowledge, MECG-E is the first model to apply time-frequency (TF) domain features to the ECG denoising task. Additionally, we conduct an in-depth comparison of the performance in two input feature types: "Complex" and "Mag.+Phase".
\item A comprehensive analysis of the composite loss function is provided, evaluating the impact of different loss functions on overall performance. We offer a detailed understanding of its contribution to model optimization.
\item We also apply power-law compression to the complex spectrum. The different STFT settings and compression exponents are then systematically compared to determine the optimal configuration.
\end{itemize}

\section{Related Work}
\label{sec:mamba}

The proposed MECG-E is built on Mamba, an advanced variant of state space models (SSMs) \cite{gu2021efficiently, gu2021combining}. SSMs are effective for sequence modeling by transforming input sequences \(x(t)\) into output sequences \(y(t)\) through hidden states \(h(t)\), which can be formulated as follows:

\begin{equation}
\begin{aligned}
\label{eq1}
h'(t) &= \mathbf{A} h(t) + \mathbf{B} x(t) \\
y(t) &= \mathbf{C} h(t)
\end{aligned}
\end{equation}
Where $\mathbf{A}$, $\mathbf{B}$, and $\mathbf{C}$ are the parameters used to define the sequence-to-sequence transformation in two stages. The ``continuous parameters" ($\Delta$, $\mathbf{A}$, $\mathbf{B}$) are transformed to ``discrete parameters" ($\mathbf{\bar{A}}$, $\mathbf{\bar{B}}$) by the zero-order hold (ZOH):

\begin{equation}
\begin{aligned}
\label{eq2}
\bar{\mathbf{A}} &= \exp(\Delta \mathbf{A}) \\
\bar{\mathbf{B}} &= (\Delta \mathbf{A})^{-1} (\exp(\Delta \mathbf{A}) - I) \cdot \Delta \mathbf{B}
\end{aligned}
\end{equation}

Traditional SSMs effectively capture temporal dependencies but are constrained by linear time-invariance, with fixed model parameters that lack adaptability to changing inputs. The Mamba model \cite{gu2023mamba} overcomes these limitations through a \textbf{selective state space mechanism} that dynamically adjusts parameters based on input. This mechanism filters irrelevant information and retains crucial data, improving performance on tasks with complex and long-range dependencies. Additionally, Mamba incorporates structural optimizations for memory usage and processing speed, particularly for GPU deployment, enabling efficient handling of long sequences.

Recently, the Mamba-based models have demonstrated success in various fields such as natural language processing \cite{lieber2024jamba}, computer vision \cite{zhu2024vision, xing2024segmamba}, and speech signal processing \cite{li2024spmamba, shams2024ssamba, chao2024investigation}.In \cite{qiang2024ecgmamba}, Mamab has been utilized as a fundamental architecture for ECG classification, showing promising performance; however, its application to ECG signal denoising, as a regression task, remains underexplored.
This paper investigates the potential of the Mamba model for denoising ECG signals, leveraging its advanced framework to effectively capture the complex dynamics of physiological data.

\section{Methodology}
\label{sec:method}
Fig.\ref{fig:ECGSE} illustrates the architecture of proposed MECG-E.
The noisy ECG signal $x \in \mathbb{R}^{1\times L}$ is first converted into the complex spectrum $X \in \mathbb{R}^{T \times F \times 2}$ by a short-time Fourier transform (STFT), where $L$, $T$ and $F$ denote the length of ECG signal, time dimensions and frequency dimensions, respectively. For the better prediction, the \textbf{power-law compression} is applied on the complex spectrum as follows:
\begin{equation}
\begin{aligned}
\label{eq3}
X^c = \left|X\right|^{c}e^{jX_p} = X_m^c e^{jX_p} = X^c_r+X^c_i
\end{aligned}
\end{equation}
where $X^c_m$, $X_p$, $X^c_r$, and $X^c_i \in \mathbb{R}^{T \times F}$ are the magnitude, wrapped phase, real, and imaginary components of the compressed spectrum, respectively. \( c \in (0,1] \) is the compression exponent, with smaller values indicating stronger compression. We compare two distinct input features to determine which yields better results for ECG reconstruction.
\begin{itemize}[leftmargin=*]
\setlength{\itemsep}{0pt}
\item \textbf{Complex Spectrum}
\begin{equation}
\begin{aligned}
\label{eq4}
X_{in,ri} = \left[X_{r}^{c}, X_{i}^{c} \right] \in \mathbb{R}^{T \times F \times 2}
\end{aligned}
\end{equation}
\item \textbf{Magnitude and Phase Spectrum}
\begin{equation}
\begin{aligned}
\label{eq5}
X_{in,mp} = \left[X_{m}^{c}, X_{p} \right] \in \mathbb{R}^{T \times F \times 2}
\end{aligned}
\end{equation}
\end{itemize}
Notably, both inputs share identical dimensions, enabling the use of a consistent model architecture. Consequently, the number of model parameters remains unchanged. The output features align with the input features. Further details can be found in Sec.\ref{sec:decoder}.

\subsection{Model Architecture}
\subsubsection{Encoder}
The encoder comprises two convolutional blocks and a dilated DenseNet \cite{pandey2020densely}. Each convolutional block includes a 2D convolutional layer, instance normalization \cite{ulyanov2016instance}, and a parametric rectified linear unit (PReLU) activation \cite{he2015delving}. The first convolutional block increases the dimensions of the input feature. The dilated DenseNet, positioned between the convolutional blocks, extends the receptive field along the time axis through multiple convolutional layers and aggregates multi-level features using dense connections. The second convolutional block downsamples the feature by expanding the stride in the convolutional layer to reduce the computational complexity. Overall, the encoder transforms the input features into a higher-dimensional TF-domain representation with a reduced sampling rate.

\subsubsection{Time-Frequency Bi-direcional Mamba (TF-Bi-Mamba)}
As illustrated in Fig.\ref{fig:ECGSE}, the TF-Bi-Mamba model consists of two sequential Bidirectional Mamba (Bi-Mamba) blocks: the Time Bi-Mamba block captures temporal dependencies, and the Frequency Bi-Mamba block captures frequency dependencies. This structure allows the model to effectively process both time and frequency aspects of the input data.

In each Bi-Mamba block, the input is processed in forward and backward directions using two parallel Mamba layers. The input for the backward Mamba is flipped before processing and reverted afterward. The outputs from both layers are concatenated, followed by a residual connection and a transposed convolution layer. This structure allows Bi-Mamba to effectively combine both current and historical information.

For the Mamba layer, we adopt the architecture outlined in \cite{gu2023mamba}, which combines the H3 block \cite{fu2023hungry} with the widely-used MLP block. In comparison to the H3 block, the Mamba layer replaces the first multiplicative gating mechanism with an activation function. Additionally, unlike the standard MLP block, the Mamba layer integrates a State-Space Model (SSM) into its main computational branch, further improving its capacity for temporal modeling. The SiLU \cite{elfwing2018sigmoid} activation function is employed to maintain a smooth and efficient transformation of input data.


\subsubsection{Decoder}
\label{sec:decoder}
We adopt the decoder architecture in \cite{lu2023mp}, which comprises two components: the Magnitude Mask Decoder and the Phase Decoder. The Magnitude Mask Decoder is designed to predict a magnitude mask $\hat{M}$, which is then multiplied with the noisy magnitude spectrum to obtain the enhanced magnitude spectrum $\hat{X}_m$:
\begin{equation}
\label{eq6}
\hat{X}_m = \left( X_m \odot \hat{M} \right)^{\frac{1}{c}}
\end{equation}
where \( \odot \) denotes element-wise multiplication and \( c \) is the compression exponent defined in Eq.\ref{eq3}. 
The Phase Decoder consists of a dilated DenseNet, a deconvolutional block, and a parallel phase estimation architecture. This architecture employs two parallel 2D convolutional layers to produce the pseudo-real $\hat{X}_p^{(r)}$ and pseudo-imaginary $\hat{X}_p^{(i)}$ components. Depending on the input features, two distinct reconstruction methods are employed.
\begin{itemize}[leftmargin=*]
\setlength{\itemsep}{0pt}
\item \textbf{Complex Spectrum:} The enhanced magnitude spectrum $\hat{X}_m$ is initially combined with the input phase $X_{p}$ to form the pseudo-complex spectrum. This spectrum is then element-wise added to the pseudo-real and pseudo-imaginary components, resulting in the final complex spectrum.
\begin{equation}
\begin{aligned}
\label{eq7}
&\hat{X}_r = \hat{X}_m\cos(X_{p})+\hat{X}_p^{(r)} \\
&\hat{X}_i = \hat{X}_m\sin(X_{p})+\hat{X}_p^{(i)} 
\end{aligned}
\end{equation}
The complex spectrum is then processed with an inverse STFT, converting it back into the reconstructed ECG signal.
\item \textbf{Magnitude and Phase Spectrum: } The pseudo-real and pseudo-imaginary components are then combined using the two-argument arctangent (Arctan2) function to predict the clean wrapped phase spectrum $\hat{X}_p$.
\begin{equation}
\label{eq8}
\hat{X}_p = \arctan \left( \frac{\hat{X}_p^{(i)}}{\hat{X}_p^{(r)}} \right) - \frac{\pi}{2} \cdot \text{Sgn}^* (\hat{X}_p^{(i)}) \cdot \left[ \text{Sgn}^* (\hat{X}_p^{(r)}) - 1 \right] \\
\end{equation}



Where $\text{Sgn}^* (t)$ equals to 1 when $t \geq 0$ and -1 otherwise. Combining the enhanced magnitude spectrum $\hat{X}_m$ with the wrapped phase spectrum $\hat{X}_p$, the ECG signal is reconstructed.
\end{itemize}

\subsection{Loss function}

The loss function is a combination of three components: time loss $\mathcal{L}_{time}$, complex loss $\mathcal{L}_{cpx}$, and consistency loss $\mathcal{L}_{con}$. 
The time and complex losses optimize the enhanced ECG in the time and TF domains, respectively. 
The consistency loss minimizes the inconsistency caused by inverse STFT and STFT.
The loss function $\mathcal{L}_{all}$ is depicted in the following equation:
\begin{equation}
\label{eq9}
\begin{aligned}
&\mathcal{L}_{all}&=&\quad \gamma_{1} \,\mathcal{L}_{time}+\gamma_{2}\,\mathcal{L}_{cpx}+\gamma_{3}\,\mathcal{L}_{con.},\\
&\mathcal{L}_{time}&=&\quad \mathbb{E}_{y,\hat{x}}\big[\lVert y-\hat{x} \rVert^{1}_{1} \big], \\
&\mathcal{L}_{cpx}&=&\quad \mathbb{E}_{Y_{c},\hat{X}_{c}}\big[\lVert Y_{c}-\hat{X}_{c} \rVert^{2}_{2} \big], \\
&\mathcal{L}_{con}&=&\quad \mathbb{E}_{\hat{x},\hat{X}_{c}}\big[\lVert \hat{X}_{c}-(\mathcal{F}_{stft}(\hat{x})\rVert^{2}_{2} \big], \\
\end{aligned}
\end{equation}
where $\gamma_{1}$, $\gamma_{2}$, and $\gamma_{3}$ are the weights for the losses, and $\hat{x}$, $y$, $\hat{X}_{c}$, and $Y_{c}$ represent the enhanced signal, target signal, and their respective complex spectrums, respectively.


\begin{table*}[!htb]
\caption{Comparison with other methods. Results marked with stars are from the referenced papers, while those without stars are reproduced by us on the same training dataset. \textbf{Bold} and \underline{underlined} values represent the best and second-best scores for the each input features, respectively.}
\label{tab:score}
\small
\vspace{0.5em}
\centering
\begin{tabular}{lcccccc}
\toprule[0.4mm]
Method & Input & SSD (au) $\downarrow$ & MAD (au) $\downarrow$ & PRD (\%) $\downarrow$  & CosSim $\uparrow$  \\ 
\midrule
\midrule
FIR filter \cite{kumar2015removal} & Waveform & 47.400 $\pm$ 88.549 & 0.677 $\pm$ 0.568 &  66.243 $\pm$ 22.137  & 0.691 $\pm$ 0.216  \\
IIR filter \cite{kumar2015removal} & Waveform & 35.728 $\pm$ 71.723 & 0.595 $\pm$ 0.531 &  61.228 $\pm$ 23.203  & 0.735 $\pm$ 0.207  \\
DRNN \cite{antczak2018deep} & Waveform &  11.492 $\pm$ 15.536 & 0.648 $\pm$ 0.390 & 118.969 $\pm$ 101.215 & 0.746 $\pm$ 0.192  \\ 
FCN-DAE \cite{chiang2019noise} & Waveform & 6.633 $\pm$ 10.885 & 0.464 $\pm$ 0.332 &  50.910 $\pm$ 26.946  & 0.877 $\pm$ 0.129 \\ 
DeepFilter \cite{romero2021deepfilter} & Waveform & 5.302 $\pm$ 8.400  & 0.395 $\pm$ 0.282 & 50.570 $\pm$ 30.703 & 0.895 $\pm$ 0.101 \\ 
DeScoD-ECG (1-shot) \cite{li2023descod} & Waveform & 4.832 $\pm$ 7.922 & 0.369 $\pm$ 0.288 &  43.588 $\pm$ 27.317 & 0.906 $\pm$ 0.110 \\ 
DeScoD-ECG (5-shot) \cite{li2023descod} & Waveform & \underline{3.915 $\pm$ 6.452} & \underline{0.334 $\pm$ 0.262} &  \underline{40.414 $\pm$ 25.227} & \underline{0.923 $\pm$ 0.090} \\ 
DeScoD-ECG (10-shot) \cite{li2023descod} & Waveform & \textbf{3.800 $\pm$ 6.227} & \textbf{0.329 $\pm$ 0.258} & \textbf{39.940 $\pm$ 25.343} & \textbf{0.926 $\pm$ 0.086} \\ 
*DeScoD-ECG (10-shot) \cite{li2023descod} & Waveform & 3.771 $\pm$ 5.713 & 0.329 $\pm$ 0.258 & 40.527 $\pm$ 26.258 & 0.926 $\pm$ 0.087 \\ 
\midrule[0.2mm]
MECG-E (Proposed) & \multirow{4}[3]{*}{Complex} & \underline{3.919 $\pm$ 8.066} & 0.326 $\pm$ 0.270 &  \textbf{37.734 $\pm$ 23.098} & \textbf{0.931 $\pm$ 0.084} \\ 
\quad -- $w/o$ time loss  &  & 7.668 $\pm$ 8.288 & 0.697 $\pm$ 0.390 & 108.094 $\pm$ 58.218 & 0.923 $\pm$ 0.076\\ 
\quad -- $w/o$ consistency loss  &  & 4.183 $\pm$ 8.725 & \textbf{0.309} $\pm$ \textbf{0.251} & \underline{38.047 $\pm$ 22.095} & 0.927 $\pm$ 0.088 \\ 
\quad -- $w/o$ complex loss  &  & \textbf{3.891 $\pm$ 7.909} & \underline{0.325 $\pm$ 0.265} & 39.714 $\pm$ 24.340 & \underline{0.930 $\pm$ 0.085} \\ 
\midrule[0.2mm]
MECG-E (Proposed) & \multirow{4}[3]{*}{Mag.+Phase} & \textbf{3.445 $\pm$ 6.493} & \underline{0.319 $\pm$ 0.252} & \underline{37.613 $\pm$ 22.389}  & \textbf{0.936 $\pm$ 0.077} \\ 
\quad -- $w/o$ time loss  &  & 7.375 $\pm$ 9.502  & 0.587 $\pm$ 0.357 & 112.942 $\pm$ 105.222 & 0.897 $\pm$ 0.090\\ 
\quad -- $w/o$ consistency loss  &  & \underline{3.611 $\pm$ 6.206} & \textbf{0.311 $\pm$ 0.235} &  38.008 $\pm$ 21.099  & \underline{0.933 $\pm$ 0.079}\\ 
\quad -- $w/o$ complex loss  &  & 3.758 $\pm$ 6.756 & 0.353 $\pm$ 0.271 &  \textbf{36.861 $\pm$ 20.694} & 0.932 $\pm$ 0.079\\ 
\bottomrule
\end{tabular}%
\end{table*}

\section{Experiment}
\label{sec:exp}

\subsection{Dataset}
To ensure consistency and comparability, we followed the dataset settings from previous studies \cite{romero2021deepfilter, li2023descod}. The clean ECG records from the QT Database\footnote{\label{note1}The QT Database and the MIT-BIH Noise Stress Test Database are available at: www.physionet.org} \cite{laguna1997database} were corrupted with noise profiles from the MIT-BIH Noise Stress Test Database\footnotemark[\getrefnumber{note1}] (NSTDB) \cite{moody1984noise}. The noise was normalized to the range of the corresponding ECG signals and rescaled by multiplying by a random factor between 0.2 and 2.0. This preprocessing pipeline ensures that the noise added to the ECG signals covers different levels of noise, allowing for a comprehensive evaluation of the denoising methods.

\subsection{Implementation Details}
The MECG-E was trained using the AdamW optimizer \cite{Loshchilov2019DecoupledWD} with a batch size of 96 and a learning rate of 1e-4 for 40 epochs. The weight decay was set to 1e-2, and an ExponentialLR scheduler \cite{li2019exponential} was employed. For the STFT, a Hamming window of 64 points and a hop length of 8 points were used. The compression exponent \( c \) in Eq. \ref{eq3} was set to 0.3. The encoder's dilated DenseNet utilized four convolutional layers with dilation sizes of \{1, 2, 4, 8\}. We configured the model with four TF-Bi-Mamba modules, each having a dimension of 32. The coefficients \(\gamma_{1}\), \(\gamma_{2}\), and \(\gamma_{3}\) in Eq. \ref{eq9} were given values of 0.5, 1, and 0.5, respectively. The implementations of MECG-E are available at the repository\footnote{https://github.com/khhungg/MECG-E}.

\subsection{Baselines and Evaluation metrics}
\label{sec:baseline}

We compared our proposed MECG-E method with several ECG denoising methods, including FIR \cite{kumar2015removal}, IIR \cite{kumar2015removal}, DRNN \cite{antczak2018deep}, FCN-DAE \cite{chiang2019noise}, DeepFilter \cite{romero2021deepfilter} and DeScoD-ECG \cite{li2023descod}. To quantitatively evaluate ECG signal distortion, we employed four distance-based metrics commonly used in previous studies \cite{romero2021deepfilter, li2023descod}: 
\begin{itemize}[leftmargin=*]
\setlength{\itemsep}{0pt}
\item \textbf{Sum of the Squares of the Distances (SSD)\cite{nygaard2001rate}:} SSD calculates the sum of squared differences between the original and processed signals, providing a detailed comparison of signal similarity over time.
\begin{equation}
\begin{aligned}
\label{eq10}
\textrm{SSD}(x,y) = \sum_{n=1}^N[y(n)-x(n)]^2. 
\end{aligned}
\end{equation}
\item \textbf{Absolute Maximum Absolute Distance (MAD)\cite{nygaard2001rate}:} MAD measures the maximum absolute difference between the original and processed signals. It is a widely used metric for evaluating ECG signal quality, focusing on the largest deviation caused by processing.
\begin{equation}
\begin{aligned}
\label{eq11}
\textrm{MAD}(x,y) = \max\left| y(n)-x(n)\right|, \quad 1 \leq n \leq N.
\end{aligned}
\end{equation}
\item \textbf{Percentage Root-Mean-Square Difference (PRD)\cite{nygaard2001rate}:} PRD is a percentage-based metric that quantifies the distortion in the processed signal. 
\begin{equation}
\begin{aligned}
\label{eq12}
\textrm{PRD}(x,y) = \sqrt{\dfrac{\sum_{n=1}^N[y(n)-x(n)]^2}{\sum_{n=1}^N[y(n)-\frac{1}{N}\sum_{n=1}^Nx(n)]}} \times 100\%.
\end{aligned}
\end{equation}
\item \textbf{Cosine Similarity (CosSim):} CosSim measures the angle between two vectors by calculating the normalized dot product based on their Euclidean L2 norms.
\begin{equation}
\begin{aligned}
\label{eq13}
\textrm{CosSim}(x,y) = \frac{\langle x,y \rangle}{\lVert x \rVert \cdot \lVert y \rVert}.
\end{aligned}
\end{equation}
\end{itemize}
For SSD, MAD, and PRD, lower values indicate better performance, while higher values are better for CosSim.

\begin{figure}[!htb]
\centerline{\includegraphics[width=1.02\columnwidth]{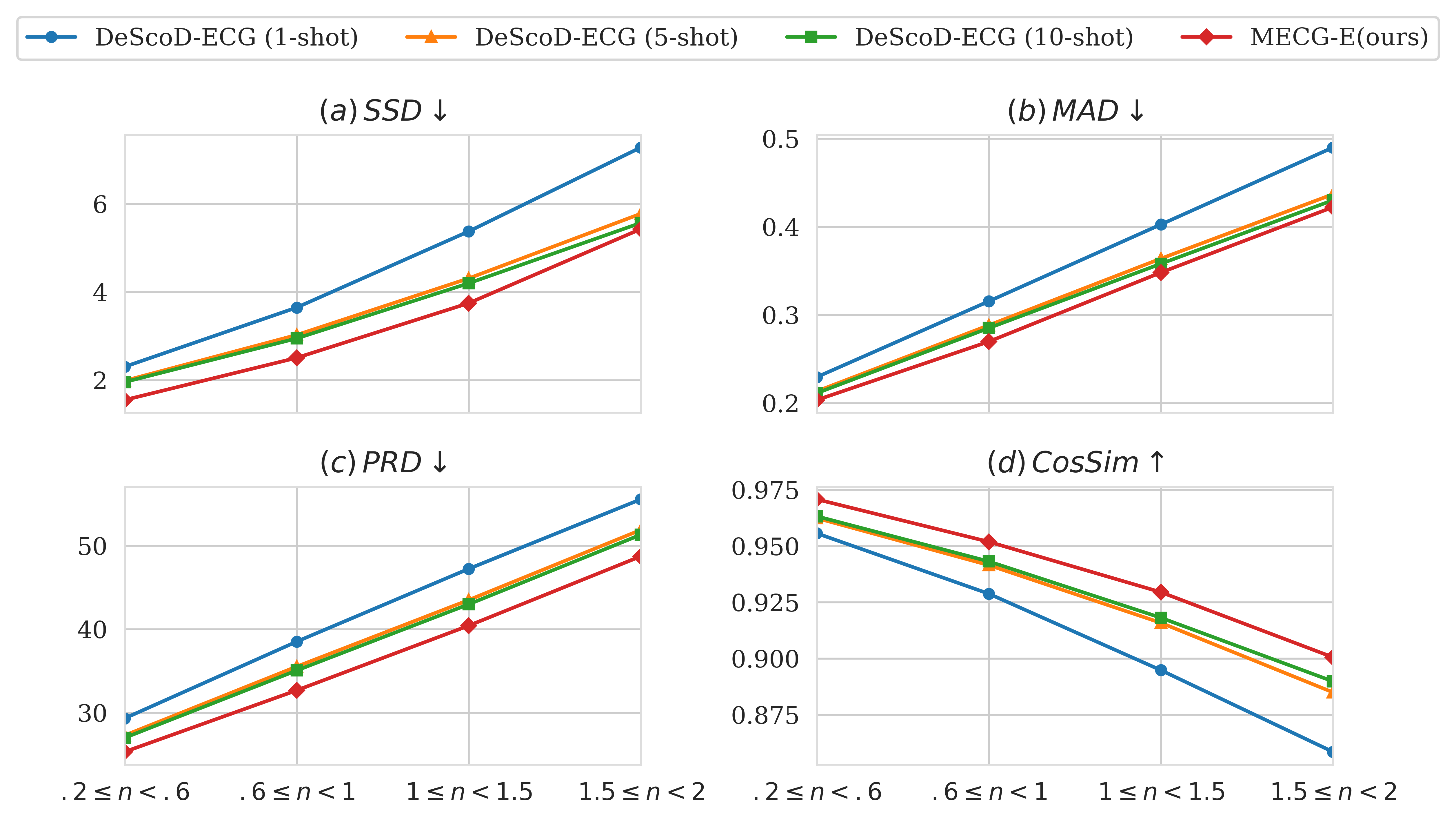}}
\caption{Evaluation metrics under different noise magnitudes. The x-axis represents the noise levels.}
\label{fig:diff_snr}
\end{figure}

\section{Results}
\label{sec:result}

The existing methods primarily use time-domain signals as input. In this paper, the proposed MECG-E is applied to the TF-domain for the first time, utilizing the 'Complex' or 'Mag.+Phase' as input features. Table \ref{tab:score} provides the mean and standard deviation of the evaluated metrics, with the bold and underlined values, respectively, representing the best and second-best results for each input feature. Among time-domain methods, the diffusion-based DeScoD-ECG outperforms other baseline approaches, especially when results are averaged across multiple runs (denoted as 'n-shots'). In the TF-domain, aside from the SSD score for 'Complex', all other metrics surpass the baseline methods, and the 'Mag.+Phase' shows slightly better performance compared to 'Complex'. Consequently, we choosed 'Mag.+Phase' as the input feature in the subsequent ablation study.

We conducted an ablation study to assess the impact of loss functions, as shown in Table \ref{tab:score}. Similar patterns were observed for both 'Complex' and 'Mag. + Phase' inputs. First, combining all three loss functions resulted in either the best or second-best performance, except for the MAD score with the 'Complex' input. Removing the time loss ($\mathcal{L}_{time}$) results in the most significant performance decrease, likely because the evaluation metrics are calculated in the time domain, making them highly sensitive to loss in the time domain. Furthermore, although adding consistency loss ($\mathcal{L}_{con}$) slightly degrades the MAD score, it significantly improved the other three metrics, demonstrating its overall benefit. In summary, removing the time loss leads to substantial degradation across all metrics, removing the consistency loss mainly impacts CosSim, and removing the complex loss most affects SSD performance. Therefore, the combination of all three loss functions provides the best balance across the evaluation metrics.


\begin{figure}[!htb]
\centerline{\includegraphics[width=1.03\columnwidth]{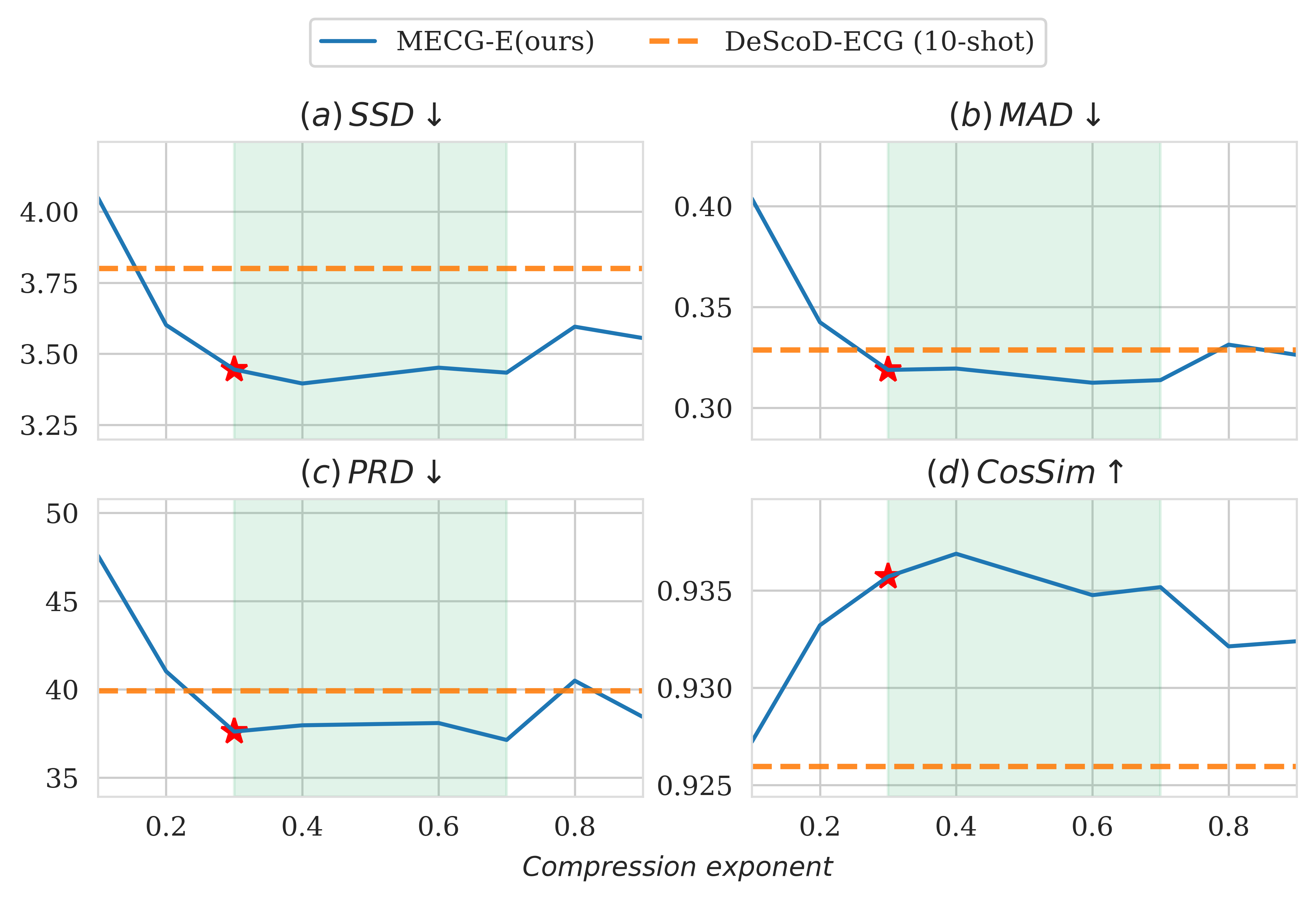}}
\caption{
Results for different compression exponents. The x-axis represents compression exponents from 0.1 to 0.9 in 0.1 increments. The red star marks the setting used in this study.
}
\label{fig:diff_com}
\end{figure}

\begin{figure}[!htb]
\centerline{\includegraphics[width=0.95\columnwidth]{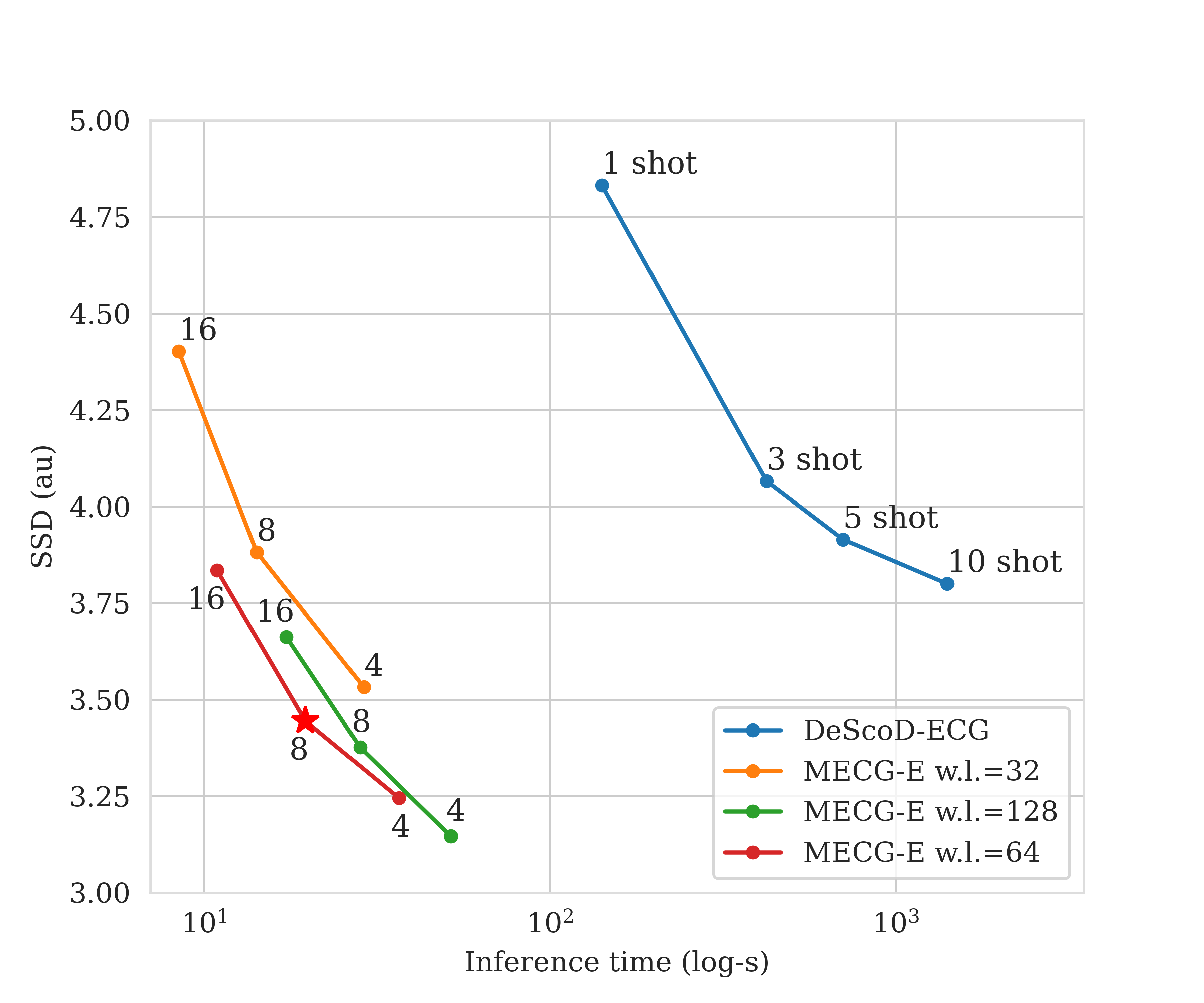}}
\caption{Relationship between SSD and inference time under various settings. The "w.l." and the number on the graph represent the window size and hop length in STFT, respectively. The red star highlights the setting used in this study.}
\label{fig:diff_time}
\end{figure}

We assessed the model's performance across various noise amplitudes, as illustrated in Fig.\ref{fig:diff_snr}, which also presents the results of DeScoD-ECG averaged over multiple runs (denoted as 'n-shots'). The results show that MECG-E (red line) consistently exceeds others with different amounts of contamination. Notably, the performance improvement of MECG-E over DeScoD-ECG at 10 shots (indicated by the gap between the green and red lines) is comparable to the improvement observed between DeScoD-ECG at 1 shot and 10 shots (the gap between the blue and green lines). This highlights the significant performance improvement achieved by MECG-E compared to the current SOTA model.

\begin{figure*}[!htb]
\centerline{\includegraphics[width=1.8\columnwidth]{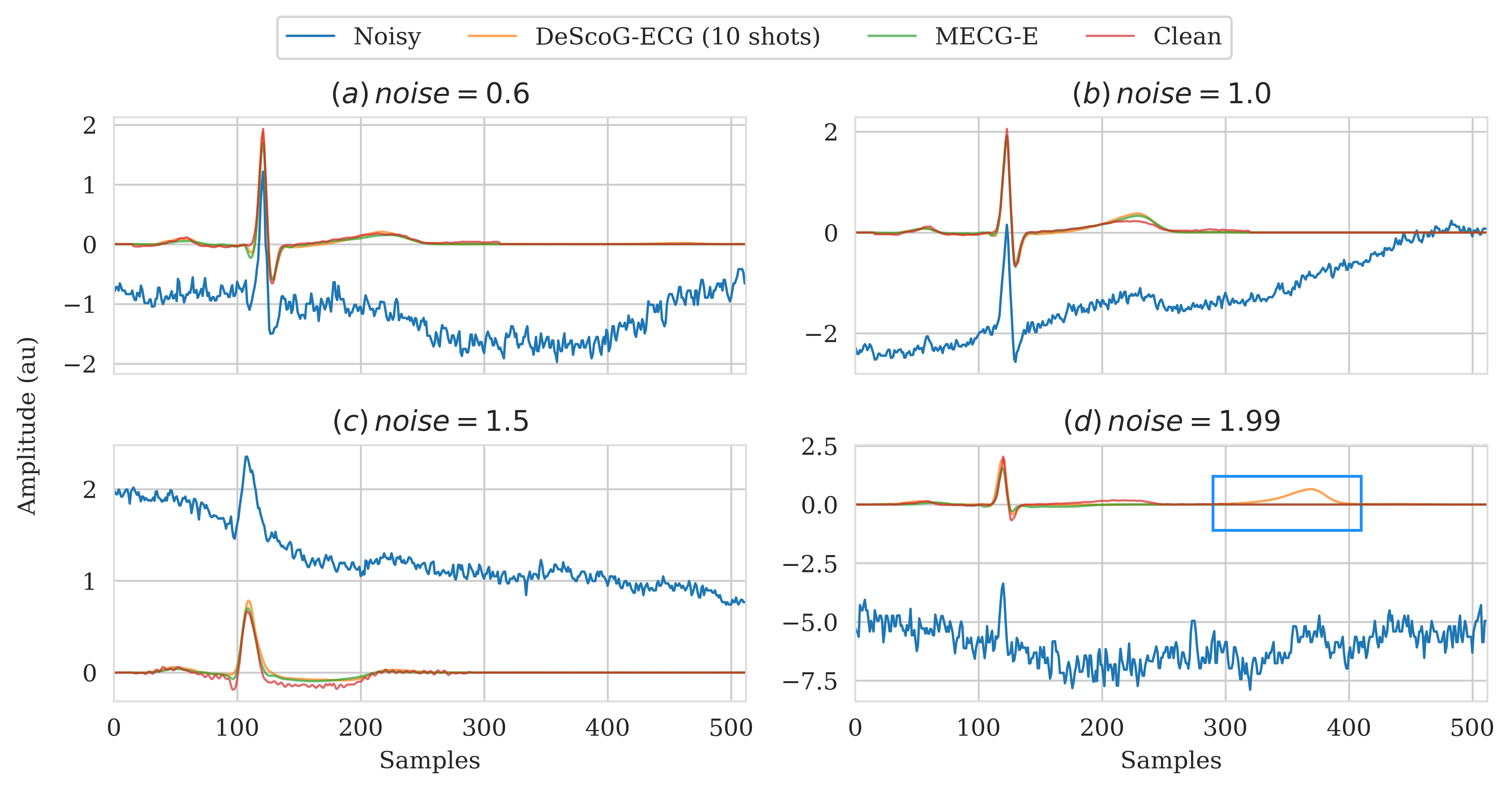}}
\caption{Reconstructed signals by the proposed MECG-E model across different noise amplitudes.
}
\label{fig:signal}
\end{figure*}

Fig.\ref{fig:diff_com} compares the effects of various compression exponents on the denoising performance. The orange dashed line is the baseline performance using DeScoD-ECG. The green-shaded area highlights where MECG-E outperforms DeScoD-ECG across all metrics. The results demonstrate that the proposed method consistently surpasses DeScoD-ECG in Cosine Similarity, and that using a compression exponent between 0.3 and 0.7 leads to better performance than the benchmark across the other three metrics. These findings underscore the importance of selecting an appropriate compression exponent to optimize performance. 

Fig.\ref{fig:diff_time} presents the analysis of inference time versus SSD across different configurations. In this plot, models located nearer to the lower-left corner, where both inference time and SSD are minimized, demonstrate superior performance. The results show that MECG-E achieves better performance compared to diffusion-based models with significantly reduced inference time. We also assessed various window sizes and hop lengths in the STFT setting. Larger window sizes and shorter hop lengths correspond to higher resolutions in the frequency and time domains, respectively. Although higher resolutions improve model performance, they also increase computational complexity and inference time. Based on our results, we conclude that a window size of 64 (indicated by the red line) provides the most efficient balance between performance and computational cost.

Fig.\ref{fig:signal} illustrates the ECG reconstruction results across various noise amplitudes. For mild noise conditions (noise = 0.6, 1.0), the reconstructed signal is closely consistent with the clean ECG signal, indicating highly accurate reconstruction. Under severe noise contamination (noise = 1.5, 1.99),  the signal morphology is highly preserved despite minor deviations between the reconstructed and clean signals. Both MECG-E and DeScoD-ECG (10 shots) produce comparable results for most signals. However, DeScoD-ECG occasionally generates artifacts in regions where no signal is present (the blue box in Fig.\ref{fig:signal}(d)), which could lead to inaccuracies in subsequent analysis. In contrast, MECG-E effectively reduces noise while preserving the integrity of the signal, making it effective across varying noise levels.

\section{Conclusion}
\label{sec:conclusion}
In this paper, we proposes MECG-E, an ECG denoising model built on the Mamba architecture. Our findings reveal that MECG-E surpasses existing methods across all evaluated metrics, even under different SNR levels. Compared to SOTA diffusion-based methods, MECG-E demonstrates greater efficiency by requiring less inference time. We also examine different STFT settings and compression exponents to identify the optimal configurations for ECG processing. In the future, we will focus on integrating MECG-E into other downstream classification tasks to further investigate its effectiveness. Additionally, exploring the model's potential for real-time applications could provide valuable insights and enhance its practical utility.


\bibliographystyle{IEEEbib}
\bibliography{strings,refs}

\end{document}